# Segment Anything Model for Brain Tumor Segmentation

Peng Zhang[a], Yaping Wang[a], *

[a]School of Electrical and Information Engineering, Zhengzhou University, Zhengzhou 450001, China;

*Corresponding author: ieypwang@zzu.edu.cn

## ABSTRACT

Brain tumor is one of the most common diseases that threatens human health. Accurate segmentation of brain tumor is important to clinical diagnosis and treatment. At present, some segmentation algorithms based on Convolutional Neural Network(CNN) and Transformer have achieved satisfactory results. These algorithms are designed for special types of datasets, and we need to design new algorithms if we want to process other different types of datasets. Therefore, there is an urgent need to find a foundation model which can adapt to different types of datasets. The Segment Anything Model(SAM), released by Meta AI, is the first foundation model in the field of image segmentation. SAM has achieved significant results in many natural image segmentation tasks. Compared with natural images, medical images have some unique properties, such as blurry boundaries, which make them more difficult to segment. Thus, it is interesting to apply SAM to the task of medical image segmentation. In this study, we evaluated the performance of SAM on brain tumor segmentation and found that: (1) SAM with box prompts performs better than that with point prompts. (2) When more points are provided, SAM's performance further improves (3) If there are too many points, the performance of SAM begins to show a decline. (4) The SAM's performance can further improve when we combine the point prompts and box prompts. (5) The performance of SAM will suffer a certain decrease if we introduce some randomness to the ideal point prompts or box prompts. (6) SAM shows remarkable performance in some special modal datasets, but is unstable in other modal datasets. (7) SAM performs well in segmenting objects with clear boundaries, but its performance is not good for objects with blurry boundaries. (8) Upon fine-tuning SAM with a subset of brain tumor images, a noticeable enhancement in its segmentation performance was observed.

**Keywords:** Segment Anything Model    brain tumor segmentation    zero-shot segmentation

## 1. INTRODUCTION

Brain tumor is a kind of malignant or benign mass that grows abnormally in the brain tissue, which may be harmful to our health. These tumors vary in type and severity, ranging from slow-growing benign tumors to malignant ones, all of which can potentially affect the structure and function of the brain[1][2].

Accurate brain tumor segmentation has some certain guidance for early clinical diagnosis and later treatment. In recent years, Convolutional Neural Network(CNN)[3] and Transformer-based[4] deep learning models have shown excellent performance in the field of computer vision. For example, some classic networks such as ResNet[5], GoogleNet[6], DenseNet[7], U-Net[8], etc. While these networks are very powerful, there are two main limitations that we must face:(1) We must have sufficient data to train these networks to meet the targets of our tasks. (2) These networks are task-special and we must redesign our network if we want to solve other problems. These limitations are particularly prominent in the field of medical image processing[9][10]. First, pixel-level annotation of images is time-consuming and labor-intensive, especially for medical images that require specialized knowledge[11][12]. Meanwhile, due to variations in imaging equipment, medical images differ significantly in terms of size, modality, resolution, and other aspects. The diversity in medical image data poses a significant challenge to the generalization capability of models[10].

Recently, large language models(LLMs) have shown satisfactory performance in the field of natural language processing(NLP). Thanks to the development of computing power and sufficient network data, many robust LLMs have been proposed, such as ChatGPT[13], GPT-4[14], LLaMA[15][16] and InterLM[17]. In response, Meta AI Research released the Segment Anything Model(SAM)[18]. As the first foundation model in the field of computer vision, it was trained on the largest image segmentation datasets(SA-1B), which contains 11 million images and over 1 billion masks. Benefiting from such sufficient training data, SAM exhibits excellent model generalization ability and supports zero-shot image segmentation with various segmentation prompts(e.g., points, boxes, and masks). From a certain perspective, SAM has learned the essence of "object". As far as we know, SAM has achieved remarkable results in natural image segmentation. However, its effectiveness in the field of medical image processing, especially in the segmentation of brain tumors, is still an area that requires further exploration. Thus, we are very curious about its performance in the task of brain tumor segmentation.

In this study, we evaluated the performance of SAM in brain tumor segmentation tasks. Our main contributions are summarized as follow:
(1) This paper comprehensively evaluates the zero-shot generalization capability of SAM in brain tumor segmentation tasks, including providing various prompting strategies for SAM, utilizing data from different imaging modalities, and segmenting different tumor regions.
(2) By adding different levels of randomness to SAM's prompting strategies, such as randomly moving point prompts or randomly scaling box prompts, we aim to better reflect SAM's performance in practical interactive segmentation scenarios.
(3) By fine-tuning SAM with a substantial amount of brain tumor datasets, it was observed that SAM's performance could be further enhanced. This highlights SAM's significant potential in downstream tasks.

## 2. METHODS

### 2.1 Brain Tumor Segmentation Datasets

We conducted experiments on the BraTS2019 datasets. The BraTS2019 datasets includes 259 cases of high-grade gliomas (HGG) and 76 cases of low-grade gliomas (LGG). Each case has four modalities: a native(T1), a post-contrast T1-weighted(T1ce), a T2-weighted(T2), and a T2 Fluid Attenuated Inversion Recovery(FLAIR). In addition to these four modalities, a manual annotation provided by experienced neuro-radiologists is also available for each case[1][19]. The annotation contains 4 classes: background(label 0), necrotic and non-enhancing tumor(label 1), peritumoral

edema(label 2) and GD-enhancing tumor(label 4). The BraTS2019 datasets includes three tumor segmentation regions: whole tumor(label 1,2 and 4), tumor core(label 1 and 4), and enhancing tumor(label 1). An example of the BraTS2019 datasets is shown in Figure 1.

Since SAM only supports 2D image inputs, and the BraTS2019 datasets is 3D MR images, we must preprocess the datasets. Firstly, we converted the BraTS2019 datasets into 2D slices along the axial plane. We got 207700 slices by slicing all the BraTS2019 datasets (including 259 cases of HGG and 76 cases of LGG, each case has four modalities) along the axial plane. Some of the slices we obtained above have no tumor areas, such as the first few slices and the last few slices of each case of data. These tumor-free slices are useless to our experiments and we should delete these slices. A straightforward and effective method is that we can make selections based on the Ground Truth(GT) of each data. If there is tumor in the GT, we retain the corresponding slices; otherwise, we remove the corresponding slices. Finally, we got our experimental datasets, which is all valid datasets containing tumors. The data preprocessing process is shown in Figure 2.

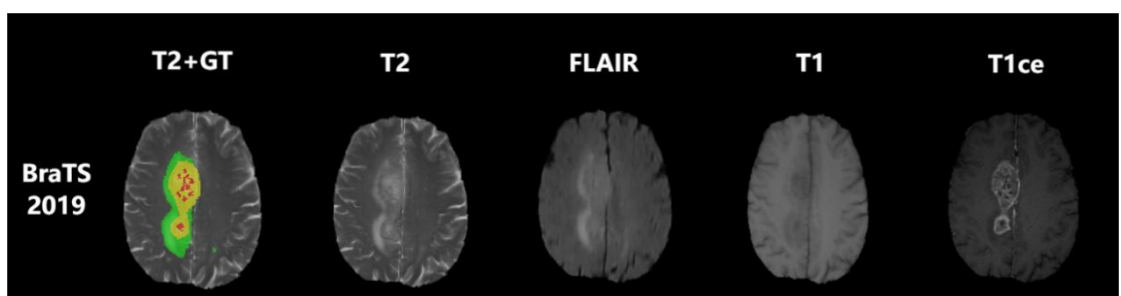

Figure 1 An example of the BraTS2019 datasets. "T2+GT" means displaying the manually annotated Ground Truth(GT) on the T2 modal image, where the green area represents peritumoral edema(label 2), the yellow area represents GD-enhancing tumor(label 4), and the red area represents necrotic and non-enhancing tumor(label 1).

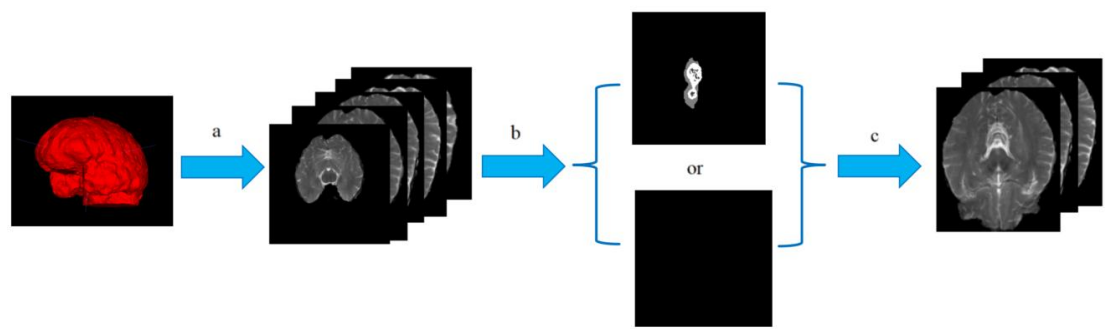

Figure 2 The display of the data preprocessing. a: convert the BraTS2019 datasets into 2D slices. b: make selections based on the Ground Truth(GT). c: get final experimental datasets.

**2.2 Brain Tumor Segmentation using SAM**

Meta AI provides the source code and online demo of SAM. For the convenience of subsequent model evaluation, we got the source code from https://segment-anything.com. We conducted six experiments:

(1) we provided SAM with 2 points, including 1 positive point and 1 negative point.
(2) we provided SAM with 10 points, including 5 positive points and 5 negative points.
(3) we provided SAM with 20 points, including 10 positive points and 10 negative points.
(4) we provided SAM with 30 points, including 15 positive points and 15 negative points.
(5) we provided SAM with 1 box.
(6) we provided SAM with 1 box and 1 positive point.

We formulated a strategy for selecting points and boxes to ensure the objectivity and accuracy of the experimental results. For positive points: (a) we first calculated the centroid of the ground truth. If the centroid is inside the ground truth, we chose it as the first positive point. (b) Next, we flattened the ground truth into a one-dimensional vector and uniformly selected other positive points. (c) If the centroid calculated in (a) is not inside the ground truth, we flattened the ground truth into a one-dimensional vector and uniformly selected all positive points. For negative points, to avoid negative points being too far from the target area, we expanded the minimum bounding box of the ground truth by 10 pixels on all sides to obtain an enlarged rectangle. Then we uniformly selected some points in the middle region between the minimum bounding box and the expanded rectangle as negative points. The details of the point prompt generation are illustrated in Algorithm 1. For box selection, we directly chose the minimum bounding box of the ground truth as the box[20].

---

**Algorithm 1** Point Prompt Generation

---

**Input:** Image **I** and its corresponding Ground Truth(GT) mask **M**, Segmentation Anything Model(**SAM**), point prompt count **N**(including N/2 positive points and N/2 negative points), function to find the smallest enclosing rectangular box of GT **F**=FindBoxPrompt().

**Step 1: If** the centroid of the ground truth is inside the ground truth:
    Chose it as the first positive point.
  **else:**
    Go to the **Step 2**.

**Step 2: If** the centroid of the ground truth is inside the ground truth:
    Flattened the ground truth into a one-dimensional vector.
    Uniformly selected (N/2 - 1) points as positive points.
  **else:**
    Flattened the ground truth into a one-dimensional vector.
    Uniformly selected N/2 points as positive points.

**Step 3:** Find the smallest enclosing rectangular box of the ground truth **MiniBox** = **F**(**M**).

**Step 4:** Expand **MiniBox** by 10 pixels on all sides to obtain an enlarged rectangle **ExpandMiniBox**.

**Step 5:** Uniformly select N/2 points in the middle region between **MiniBox** and **ExpandMiniBox**.

---

## 2.2 Evaluation Metrics

To comprehensively evaluate the segmentation performance of SAM, we employed three common evaluation metrics:

(1) Dice coefficient(Dice)[21]: Dice is widely used to evaluate the performance of image segmentation models. It measures the degree of overlap between the predicted and ground truth labels. The Dice score ranges from 0 to 1, with higher values indicating better overlap between the predicted and ground truth labels. Assuming that the ground truth of manual segmentation is G and the network predicted segmentation result is P, then the calculation formula of Dice is:

$$Dice = \frac{2|G \cap P|}{|G| + |P|}$$

(2) Hausdorff Distance(HD)[22]: HD measures the performance of segmentation models by computing the maximum mismatch between two sets of contour points, which can reflect the distance between each point in the prediction to the point in the ground truth. Assuming that A and B represent a set of point sets of manual segmentation results and algorithm segmentation results respectively, the calculation formula of HD is:

$$HD = \max(h(A,B), h(B,A))$$

$$h(A,B) = \max_{a \in A} \left\{ \min_{b \in B} \|a - b\| \right\}$$

$$h(B,A) = \max_{b \in B} \left\{ \min_{a \in A} \|b - a\| \right\}$$

(3) Average Symmetric Surface Distance(ASSD)[23]: ASSD is used to measure the degree of surface alignment in segmentation results. It computes the minimum distance between each point on the segmentation surface and the ground truth surface, as well as the minimum distance between each point on the ground truth surface and the segmentation boundary. The average of these two types of minimum distances is taken, with smaller values indicating better segmentation performance. The calculation formula of ASSD is:

$$ASSD = \frac{1}{2} \left( \frac{\sum_{x \in X} \min_{y \in Y} d(x,y)}{len(X)} + \frac{\sum_{y \in Y} \min_{x \in X} d(y,x)}{len(Y)} \right)$$

X is the set of boundary contour pixels from manual segmentation results, Y is the set of boundary contour pixels from algorithmic segmentation results, and d(x,y) represents the Euclidean distance between a boundary contour pixel x from manual segmentation results and a boundary contour pixel y from algorithmic segmentation results.

## 3. RESULTS

### 3.1 Quantitative evaluation

The brain tumor segmentation performed by SAM was evaluated across the aforementioned metrics. The quantitative evaluation of various prompting types is shown in Table 1.

Table 1. Segmentation performance of SAM on BraTS2019 datasets with different prompts.

| Region | Method | Prompts | T1 | | | T1ce | | | T2 | | | FLAIR | | |
|---|---|---|---|---|---|---|---|---|---|---|---|---|---|---|
| | | | Dice ↑ | HD ↓ | ASSD ↓ | Dice ↑ | HD ↓ | ASSD ↓ | Dice ↑ | HD ↓ | ASSD ↓ | Dice ↑ | HD ↓ | ASSD ↓ |
| Whole tumor | SAM | 2 points | 0.4174 | 27.9233 | 7.9372 | 0.4194 | 24.7791 | 7.1544 | 0.4761 | 25.8699 | 6.8054 | 0.4209 | 24.6662 | 7.0813 |
| | SAM | 10 points | 0.6553 | 20.6195 | 2.5763 | 0.6723 | 18.8586 | 2.0258 | 0.7354 | 19.2708 | 1.3135 | 0.6719 | 18.8708 | 2.0333 |
| | SAM | 20 points | 0.6135 | 22.5458 | 3.1381 | 0.6522 | 20.7484 | 2.2640 | 0.7011 | 20.8633 | 1.5488 | 0.6519 | 20.7485 | 2.2765 |
| | SAM | 30 points | 0.6301 | 21.7602 | 2.9798 | 0.6516 | 19.8056 | 2.3108 | 0.7077 | 20.1296 | 1.4963 | 0.6516 | 19.8056 | 2.3108 |
| | SAM | 1 box | 0.6989 | 17.2557 | 1.1799 | 0.6564 | 15.8840 | 1.4546 | 0.7535 | 16.9590 | 0.9094 | 0.6564 | 15.8840 | 1.4546 |
| | SAM | 1 box + 1point | 0.7032 | 18.0177 | 1.2269 | 0.6717 | 16.7765 | 1.4427 | 0.7661 | 18.1716 | 0.8792 | 0.6717 | 16.7765 | 1.4427 |
| Tumor Core | SAM | 2 points | 0.4294 | 22.9007 | 7.5402 | 0.5887 | 18.7482 | 4.7736 | 0.4941 | 21.0223 | 6.1563 | 0.5889 | 18.8427 | 4.7875 |
| | SAM | 10 points | 0.6606 | 16.5422 | 2.4385 | 0.7711 | 15.0163 | 1.4097 | 0.7234 | 15.6083 | 1.2741 | 0.7715 | 14.9999 | 1.3922 |
| | SAM | 20 points | 0.6141 | 18.2911 | 2.8972 | 0.7489 | 15.9774 | 1.5605 | 0.6825 | 17.0129 | 1.4908 | 0.7487 | 15.9818 | 1.5594 |
| | SAM | 30 points | 0.6298 | 17.7419 | 2.7798 | 0.7509 | 15.6975 | 1.5973 | 0.6887 | 16.6573 | 1.4585 | 0.7509 | 15.6975 | 1.5973 |
| | SAM | 1 box | 0.7285 | 13.3363 | 0.8939 | 0.7823 | 12.7772 | 0.7290 | 0.7605 | 13.2337 | 0.7529 | 0.7823 | 12.7772 | 0.7290 |
| | SAM | 1 box + 1point | 0.7460 | 14.1889 | 0.8751 | 0.8029 | 13.6031 | 0.6878 | 0.7826 | 14.2799 | 0.6962 | 0.8029 | 13.6031 | 0.6878 |
| Enhancing tumor | SAM | 2 points | 0.3311 | 22.4959 | 7.9852 | 0.5068 | 18.2667 | 5.0903 | 0.3925 | 20.5919 | 6.5964 | 0.5075 | 18.2012 | 5.0929 |
| | SAM | 10 points | 0.5299 | 16.4740 | 2.7836 | 0.6514 | 14.8369 | 1.6812 | 0.5942 | 15.5222 | 1.5908 | 0.6507 | 14.8695 | 1.7081 |
| | SAM | 20 points | 0.4912 | 18.4377 | 3.2198 | 0.6275 | 15.9732 | 1.8499 | 0.5576 | 17.1426 | 1.7782 | 0.6277 | 15.9718 | 1.8517 |
| | SAM | 30 points | 0.5040 | 17.9003 | 3.1245 | 0.6312 | 15.7195 | 1.8869 | 0.5638 | 16.8063 | 1.7439 | 0.6312 | 15.7195 | 1.8869 |
| | SAM | 1 box | 0.5916 | 12.5418 | 1.2538 | 0.6717 | 12.1618 | 0.9898 | 0.6278 | 12.6252 | 1.0881 | 0.6717 | 12.1618 | 0.9898 |
| | SAM | 1 box + 1point | 0.6035 | 13.5729 | 1.2506 | 0.6796 | 13.2989 | 0.9823 | 0.6442 | 13.8973 | 1.0435 | 0.6796 | 13.2989 | 0.9823 |

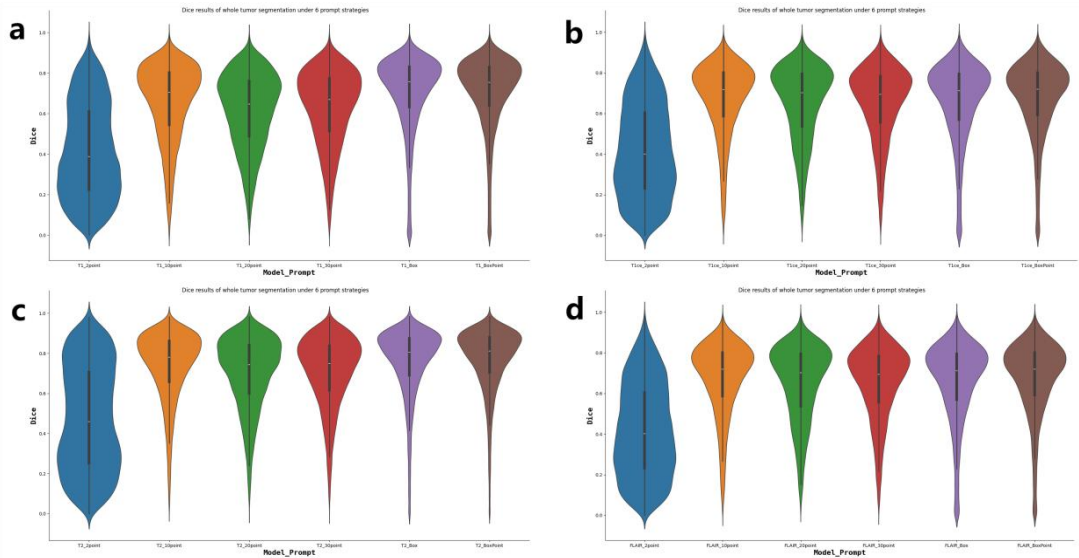

Figure 3 Dice results of the whole tumor segmentation under different prompting strategies. (a)T1 (b)T1ce (c)T2 (d)FLAIR. The abscissa is the abbreviation of different segmentation strategies in different modalities. For example, "T1_Box" represents the segmentation result of the T1 modal image with 1 box prompt.

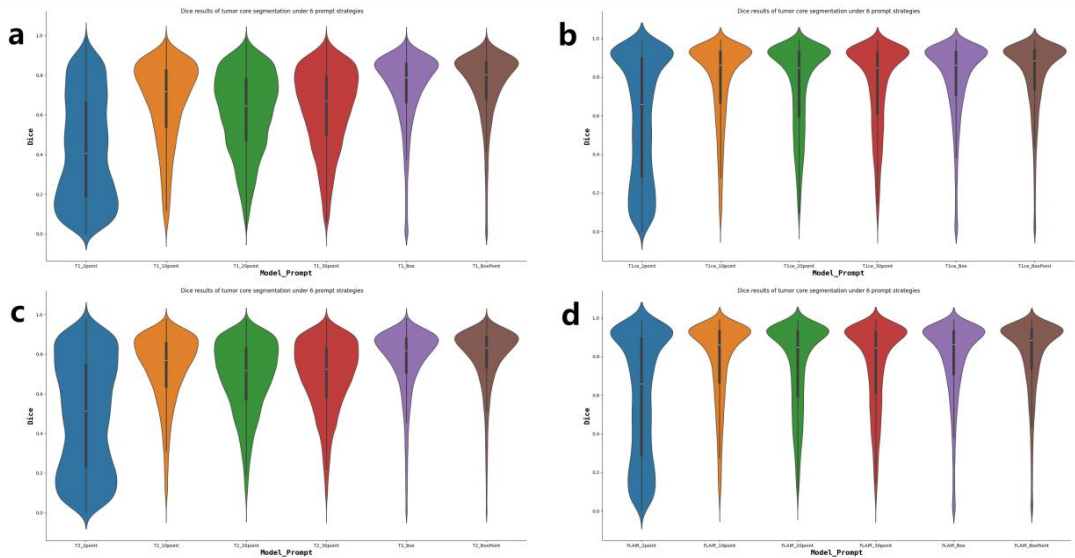

Figure 4　Dice results of the tumor core segmentation under different prompting strategies. (a)T1 (b)T1ce (c)T2 (d)FLAIR. The abscissa is the abbreviation of different segmentation strategies in different modalities. For example, "T1_Box" represents the segmentation result of the T1 modal image with 1 box prompt.

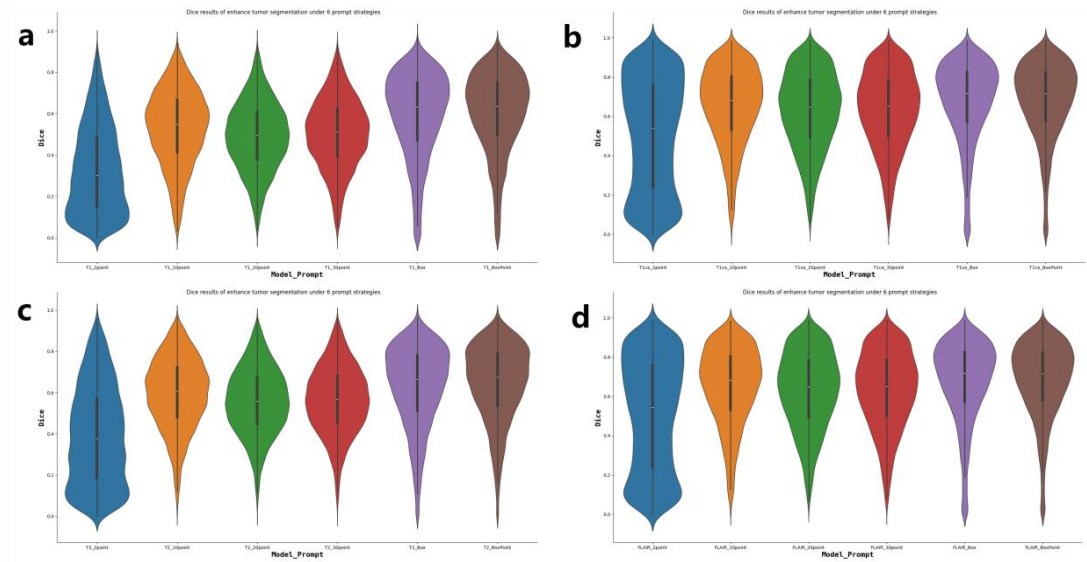

Figure 5　Dice results of the enhancing tumor segmentation under different prompting strategies. (a)T1 (b)T1ce (c)T2 (d)FLAIR. The abscissa is the abbreviation of different segmentation strategies in different modalities. For example, "T1_Box" represents the segmentation result of the T1 modal image with 1 box prompt.

From the experimental results in Table 1 and Figures 3 to 6, we can draw the following conclusions:
(1) Segmentation performance under different prompts
For the same segmentation area under the same imaging modality, the segmentation performance varies significantly with different prompting mechanisms. Firstly, we can observe from Table 1 that SAM's performance with 10 point prompts is superior to that with 2 point prompts(For example, in the task of the whole tumor segmentation of T2 modality data, SAM with 2 point prompts achieves a Dice value of 0.476, while SAM with 10 point prompts achieves a Dice value of 0.7354). In the violin plots from Figure 3 to Figure 5, it is also evident that the segmentation performance of SAM based on 2 point prompts is highly unstable, with a large fluctuation range in Dice values. On the other hand, the segmentation performance of SAM based on 10 point prompts is relatively stable, with a smaller variation range in Dice values. This indicates that 2 points are insufficient to provide SAM with adequate prompt messages for segmentation, while 10 points can offer more information, leading to a relative improvement in SAM's segmentation performance. This indicates that appropriately increasing the number of point prompt can enhancing SAM's segmentation performance. This can be explained as: If we provide SAM with more point prompts, we can impose certain constraints on SAM's segmentation. For example, if we uniformly add a certain number of negative points around the segmentation area, we can reduce some false positive areas to benefit SAM's performance.

However, as the number of point prompts continues to increase, we can also observe that the performance of SAM with 20 point prompts and that with 30 point prompts begins to show a decline(For example, in the tasks of the whole tumor and enhancing tumor segmentation, the segmentation performance of SAM based on 20 point prompts and 30 point prompts starts to decline. Specifically, in the T1 modality of whole tumor segmentation, we can find that compared to the Dice value of 0.6553 when using 10 point prompts, the segmentation performance of SAM drops to Dice values of 0.6135 and 0.6301 when using 20 and 30 point prompts, respectively). From the segmentation results in Figure 6, we can also observe this result. When we increase the number of point prompt to 20 or 30, we can notice an increasing of the false positive regions. For instance, in the whole tumor segmentation of T2 modality data in Figure 6, SAM based on 20 and 30 point prompts exhibits a significant number of false-positive segmentation results in the upper right corner of the brain tumor. This is primarily because when we have too many point prompts, such as 20 or 30 point prompts, many point prompts tend to be relatively close to each other. Additionally, the effective range (or "receptive field") of each point prompt is relatively limited. Consequently, many point prompts cannot exert their intended performance. Moreover, when there are too many point prompts, many of them will be located at the boundary of the segmentation area. As the boundary of brain tumors is generally fuzzy, too many point prompts will lead to an increasing of the false positive areas so that it will lead to a decrease in segmentation performance.

In summary, appropriately increasing the number of prompt points will bring some segmentation constraints to SAM, which is beneficial to improve the segmentation performance. However, too many point prompts may have a counterproductive effect, not only rendering many point prompts ineffective but also increasing false positive areas, ultimately leading to lower segmentation performance.

In addition to point prompt, SAM also supports box prompt. Given a box prompt, the region inside the box prompt is the area SAM aims to segment. From the quantitative segmentation results in Table 1, we can observe that the segmentation performance of SAM based on 1 box prompt is superior to that based on point prompt. For example, in the segmentation of the tumor core of T2 modality data, SAM based on 10 point prompts achieved the best segmentation, yielding a Dice value of 0.7234. In contrast, SAM based on 1 box prompt achieves a Dice value of 0.7605 under the same conditions. Similarly, for the segmentation of the tumor core of T1 modality data, SAM based on 10 point prompts achieves a Dice value of 0.6606, while SAM based on 1 box prompt achieves a Dice value of 0.7285. This can be explained as follows: compared with natural images, the boundary of tumor is often blurry, making it easy to produce false positive areas. Box prompt can effectively confine the segmentation results within the box, reducing the false positive areas and improving SAM's segmentation performance. In the violin plots (Figure 3 to Figure 5), we can also observe that the segmentation results of SAM based on 1 box prompt are relatively concentrated, and the Dice values are relatively higher. This indicates that the box prompt can provide SAM with accurate segmentation boundaries, thereby stabilizing the segmentation performance of SAM.

When we combine the point and box prompt, we can observe that the segmentation performance of SAM based on 1 positive point and 1 box prompts is superior to SAM based only 1 box prompt. For instance, in the segmentation of the tumor core of T1ce modality data, SAM based on 1 box prompt achieves a segmentation Dice value of 0.7823, while SAM based on 1 positive point and 1 box prompts achieves a Dice value of 0.8029. This can be explained as follows: Box prompt provides SAM with accurate segmentation boundaries, reducing false positive areas. If we add 1 positive point to the box prompt, the additional positive point further assist SAM in refining the segmentation position, thereby enhancing SAM's segmentation performance.

(2) Segmentation performance under different medical image modalities

Under the same segmentation region and prompting strategy, SAM's segmentation performance also exhibits significant differences when using different modality data. For instance, in the task of the whole tumor segmentation with 10 point prompts, among the four imaging modality data, the segmentation performance of T1 modality data is the lowest with a Dice value of 0.6553, while T2 modality data has the best segmentation performance with a Dice value of 0.7354. In the violin plots from Figure 3 to Figure 5, we can also observe that the Dice values across the three modalities data, T1ce, T2, and FLAIR, are generally higher than those of the T1 modality data. Specifically, in the tumor core segmentation in Figure 4, it is apparent that the Dice values for segmentation using T1ce and FLAIR modality data are higher than that using the T1 modality data. This can be explained from Figure 1: we can observe that the segmentation boundaries of T1 modality data is the most blurred among four imaging modalities, leading to relatively poor segmentation results for T1 modality data. On the other hand, the segmentation boundaries for T2, T1ce, and FLAIR modality data are relatively clear, resulting in better segmentation performance. However, it is essential to note that while the boundaries are clear in T1ce modality data, they mainly delineate the enhancing tumor and tumor core. In contrast, the outermost part, peritumoral edema, exhibits significant blurriness. This leads to poor segmentation of the whole tumor in the T1ce modality data because the peritumoral edema which is located in the whole tumor is almost indistinguishable. In contrast, the segmentation of the tumor core and enhancing tumor is better in T1ce modality data because the boundaries of the enhancing tumor and tumor core are clearer.

(3) Segmentation performance under different segmentation areas

From the experimental results in Table 1, we can observe that the segmentation performance of the tumor core is the best overall, followed by the whole tumor, while the segmentation performance of the enhancing tumor is relatively weaker. The tumor core comprises the necrotic and non-enhancing tumor as well as the GD-enhancing tumor, the enhancing tumor includes the necrotic and non-enhancing tumor, and the whole tumor encompasses the necrotic and non-enhancing tumor, the peritumoral edema, and the GD-enhancing tumor. As shown in Figure 1, the GD-enhancing tumor (yellow area) is relatively clear, while the peritumoral edema (green area) is blurred, and the necrotic and non-enhancing tumor (red area) is not very clear due to its small size. This explains why, in terms of segmentation performance across the three tumor regions, the tumor core achieves the best segmentation, with the enhancing tumor exhibiting relatively weaker segmentation results. From the segmentation results in Figure 6, we can also observe that due to the clear boundary contours in GD-enhancing tumor, the segmentation results for the tumor core are relatively good. Specifically, in the segmentation of the tumor core of the T1ce modality data, we can see that the segmentation results under the six prompting strategies are generally similar to the Ground Truth. In contrast, the boundary contours of peritumoral edema are very blurry compared to GD-enhancing tumor, making accurate segmentation challenging. In the whole tumor segmentation result in Figure 6, we can observe that due to the blurry boundaries of peritumoral edema on the outer periphery of the tumor, there are some false positive regions in the segmentation results.

**3.2 Visual Comparison**

In order to show the effect of SAM on brain tumor segmentation more intuitively, we show the segmentation performance of SAM on the whole tumor, tumor core, and enhancing tumor under different prompting mechanisms, as shown in Figure 6 below.

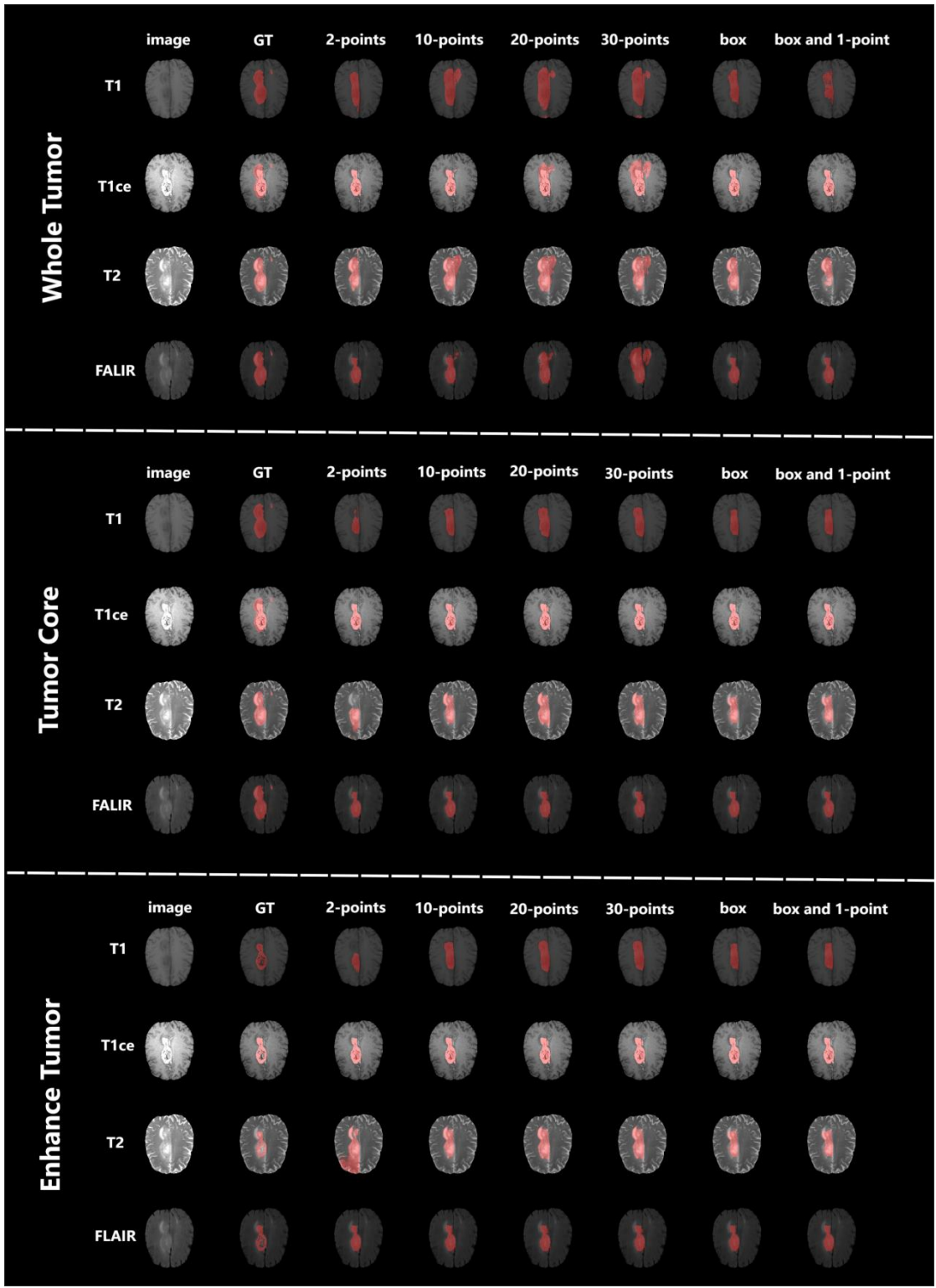

Figure 6  Visualization results of SAM segmentation for the brain tumor under different prompts.

# 4. DISCUSSION

## 4.1 Further experiments

### 4.1.1 Randomness

In the above experiments, both the point and box prompts were selected based on the Ground Truth (GT). Specifically, positive points were chosen uniformly distributed within the GT, and negative points were selected uniformly distributed around the boundary of the GT. The bounding box was chosen as the minimum bounding rectangle of the GT. The primary purpose of this approach is to select the ideal point and box prompt to reflect the best performance of SAM.

However, in real-world interactive segmentation applications, we cannot provide SAM with the most ideal point and box prompts. Therefore, in the subsequent experiments, we add some randomness to the prompts gained from the Algorithm 1. Specifically, we randomly applied scaling/moving operations to the box/point obtained from Algorithm 1, simulating some of the randomness that may exist in real interactive segmentation scenarios. For the box obtained from Algorithm 1, we randomly scaled them by 0 to 5 pixels, 5 to 10 pixels, and 10 to 15 pixels. For the point obtained from Algorithm 1, we randomly moved them by 0 to 5 pixels, 5 to 10 pixels, and 10 to 15 pixels. This allows us to observe SAM's segmentation performance in real interactive segmentation applications with some level of randomness. The quantitative evaluation of various prompting types is shown in Table 2.

Table 2. Results of SAM on BraTS2019 datasets. Dice, HD, and ASSD are employed for evaluation with SAM.

| Region | Shift | Prompts | T1 | | | T1ce | | | T2 | | | FLAIR | | |
|---|---|---|---|---|---|---|---|---|---|---|---|---|---|---|
| | | | ΔDice | ΔHD | ΔASSD | ΔDice | ΔHD | ΔASSD | ΔDice | ΔHD | ΔASSD | ΔDice | ΔHD | ΔASSD |
| Whole tumor | 0-5 | 2 points | -0.0066 | 0.1886 | 0.8696 | -0.0066 | 0.1367 | 0.2385 | -0.0119 | 0.1213 | 0.3054 | -0.0088 | -4.5991 | 0.3256 |
| | | 10 points | -0.0118 | 0.0455 | 0.1953 | -0.0125 | 0.0876 | 0.1709 | -0.0112 | 0.0738 | 0.1468 | -0.0113 | 0.0189 | 0.1589 |
| | | 20 points | -0.0105 | 0.1392 | 0.1941 | -0.0111 | 0.1098 | 0.1705 | -0.0108 | 0.1281 | 0.1278 | -0.01 | 0.1113 | 0.1404 |
| | | 30 points | -0.0062 | 0.0545 | 0.1021 | -0.0053 | 0.0943 | 0.046 | -0.0069 | 0.105 | 0.0714 | -0.0061 | 0.1016 | 0.0615 |
| | | 1 box | -0.0329 | 0.9029 | 0.1856 | -0.0216 | 0.7071 | 0.1046 | -0.0314 | 0.8531 | 0.1489 | -0.0212 | 0.7044 | 0.1057 |
| | | 1 box + 1point | -0.0353 | 4.3624 | 0.1563 | -0.0045 | 4.263 | -0.0229 | -0.0521 | 2.9241 | 0.2557 | -0.0063 | 4.3007 | 0.0014 |
| | 5-10 | 2 points | -0.0455 | 1.0199 | 1.8346 | -0.0486 | 2.292 | 2.0929 | -0.0747 | 1.7822 | 2.4092 | -0.0484 | 2.5441 | 2.1859 |
| | | 10 points | -0.1162 | 0.5929 | 2.8716 | -0.1117 | 0.5792 | 2.2311 | -0.1126 | 0.7049 | 2.1593 | -0.1135 | 0.6308 | 2.3255 |
| | | 20 points | -0.0729 | 0.5071 | 1.4874 | -0.0811 | 0.4637 | 1.3954 | -0.0758 | 0.6331 | 1.1923 | -0.0797 | 0.4879 | 1.3655 |
| | | 30 points | -0.0684 | 0.3807 | 1.5268 | -0.068 | 0.4245 | 1.2785 | -0.065 | 0.5404 | 1.1393 | -0.0683 | 0.4282 | 1.267 |
| | | 1 box | -0.1002 | 2.6431 | 0.7057 | -0.068 | 2.2114 | 0.4886 | -0.1033 | 2.5119 | 0.6349 | -0.0684 | 2.2219 | 0.4942 |
| | | 1 box + 1point | -0.1383 | 6.8522 | 1.2016 | -0.1001 | 6.9812 | 0.9111 | -0.1571 | 5.3789 | 1.165 | -0.0989 | 6.9388 | 0.8905 |
| | 10-15 | 2 points | -0.1141 | 5.1672 | 5.4077 | -0.1228 | 7.5402 | 6.0819 | -0.1644 | 6.341 | 6.4357 | -0.1242 | 7.8946 | 6.3048 |
| | | 10 points | -0.2892 | 2.9117 | 8.7605 | -0.2891 | 3.1257 | 7.7675 | -0.3059 | 3.3732 | 7.6773 | -0.2892 | 3.1227 | 7.7875 |
| | | 20 points | -0.2136 | 0.9828 | 5.6322 | -0.2297 | 0.9399 | 5.1548 | -0.2276 | 1.3895 | 4.7496 | -0.2295 | 1.0429 | 5.157 |
| | | 30 points | -0.2097 | 0.9704 | 5.5517 | -0.2151 | 1.2351 | 5.0508 | -0.2146 | 1.4931 | 4.7646 | -0.2164 | 1.2686 | 5.099 |
| | | 1 box | -0.1639 | 4.0803 | 1.4562 | -0.1185 | 3.7037 | 1.131 | -0.1721 | 3.9872 | 1.3232 | -0.1195 | 3.7306 | 1.136 |

| Region | Range | Prompt | | | | | | | | | | | |
|---|---|---|---|---|---|---|---|---|---|---|---|---|---|
| | | 1 box + 1point | -0.2411 | 9.0467 | 2.9478 | -0.2022 | 9.408 | 2.5536 | -0.2719 | 7.7168 | 2.8528 | -0.1998 | 9.4686 | 2.4958 |
| | 0-5 | 2 points | -0.0123 | 0.3368 | 0.3825 | -0.0258 | 0.7917 | 0.6775 | -0.0219 | 0.4972 | 0.6116 | -0.0284 | 0.7422 | 0.6971 |
| | | 10 points | -0.0156 | 0.1354 | 0.2124 | -0.0142 | 0.1298 | 0.1132 | -0.0156 | 0.1663 | 0.176 | -0.0136 | 0.1543 | 0.1015 |
| | | 20 points | -0.0181 | 0.2685 | 0.3012 | -0.0176 | 0.26 | 0.1632 | -0.0169 | 0.2323 | 0.1833 | -0.0179 | 0.2536 | 0.174 |
| | | 30 points | -0.0132 | 0.1824 | 0.185 | -0.0131 | 0.1813 | 0.0615 | -0.0119 | 0.1217 | 0.1185 | -0.0148 | 0.2174 | 0.1184 |
| | | 1 box | -0.0467 | 0.8881 | 0.2234 | -0.0261 | 0.4322 | 0.1374 | -0.0449 | 0.857 | 0.1943 | -0.0263 | 0.4475 | 0.1388 |
| | | 1 box + 1point | -0.0559 | 3.2307 | 0.2282 | -0.0244 | 2.1098 | 0.0713 | -0.063 | 2.3603 | 0.2648 | -0.0243 | 2.1132 | 0.0676 |
| Tumor Core | 5-10 | 2 points | -0.0793 | 2.5456 | 2.7551 | -0.1438 | 4.6295 | 4.0937 | -0.1094 | 3.1345 | 3.3695 | -0.1463 | 4.7619 | 4.253 |
| | | 10 points | -0.1545 | 1.5428 | 3.4429 | -0.1639 | 1.4832 | 2.6814 | -0.1497 | 1.4773 | 2.6151 | -0.1602 | 1.4444 | 2.5801 |
| | | 20 points | -0.1151 | 1.3257 | 2.1838 | -0.1349 | 1.2787 | 1.6747 | -0.1126 | 1.2745 | 1.5873 | -0.1323 | 1.297 | 1.6348 |
| | | 30 points | -0.1178 | 1.1976 | 2.477 | -0.125 | 1.1651 | 1.7883 | -0.1053 | 1.0499 | 1.7063 | -0.1231 | 1.1843 | 1.7651 |
| | | 1 box | -0.1533 | 2.7702 | 0.961 | -0.0975 | 1.5869 | 0.6066 | -0.1538 | 2.8205 | 0.8596 | -0.0963 | 1.616 | 0.5929 |
| | | 1 box + 1point | -0.1956 | 6.1923 | 1.3447 | -0.1524 | 4.5157 | 0.9634 | -0.2023 | 5.1368 | 1.302 | -0.1512 | 4.4937 | 0.9619 |
| | 10-15 | 2 points | -0.1804 | 8.0482 | 7.7334 | -0.2965 | 11.7977 | 10.0748 | -0.2349 | 8.8295 | 8.6707 | -0.2948 | 11.3911 | 9.8758 |
| | | 10 points | -0.3564 | 5.4217 | 10.4376 | -0.4309 | 5.3764 | 9.6499 | -0.3755 | 4.9912 | 9.0162 | -0.4302 | 5.359 | 9.7167 |
| | | 20 points | -0.2902 | 2.9235 | 7.5419 | -0.3912 | 3.4568 | 7.4686 | -0.305 | 2.8531 | 6.0852 | -0.391 | 3.5044 | 7.4951 |
| | | 30 points | -0.2968 | 3.2213 | 8.0858 | -0.3754 | 3.4043 | 7.4828 | -0.2978 | 2.8303 | 6.2623 | -0.3764 | 3.4615 | 7.5438 |
| | | 1 box | -0.2475 | 4.5753 | 1.9858 | -0.1904 | 2.9701 | 1.4453 | -0.2485 | 4.6619 | 1.7689 | -0.1905 | 2.9845 | 1.4437 |
| | | 1 box + 1point | -0.3303 | 8.8 | 3.2691 | -0.3253 | 7.7003 | 2.8654 | -0.3372 | 7.7464 | 3.0416 | -0.3258 | 7.7719 | 2.8555 |
| | 0-5 | 2 points | -0.0139 | 0.1148 | 0.4119 | -0.0408 | 0.7125 | 0.8553 | -0.021 | 0.355 | 0.6377 | -0.0396 | 0.7984 | 0.8518 |
| | | 10 points | -0.0157 | 0.0817 | 0.2523 | -0.0174 | 0.2745 | 0.1651 | -0.0163 | 0.1936 | 0.2047 | -0.0169 | 0.2137 | 0.1254 |
| | | 20 points | -0.0175 | 0.2714 | 0.347 | -0.0214 | 0.3737 | 0.1714 | -0.0157 | 0.221 | 0.1836 | -0.0214 | 0.3682 | 0.1981 |
| | | 30 points | -0.0131 | 0.1516 | 0.2342 | -0.0178 | 0.2802 | 0.114 | -0.0126 | 0.1492 | 0.1439 | -0.018 | 0.3153 | 0.1125 |
| | | 1 box | -0.0382 | 1.0876 | 0.1771 | -0.0243 | 0.5934 | 0.1188 | -0.0372 | 1.1107 | 0.1488 | -0.0238 | 0.5953 | 0.1165 |
| | | 1 box + 1point | -0.0388 | 3.8975 | 0.1216 | -0.0232 | 2.4047 | 0.0392 | -0.0508 | 2.7665 | 0.1954 | -0.0226 | 2.3922 | 0.0384 |
| Enhancing tumor | 5-10 | 2 points | -0.0735 | 2.3537 | 3.0687 | -0.1656 | 4.5292 | 4.7445 | -0.1033 | 3.104 | 3.7323 | -0.1641 | 4.5887 | 4.639 |
| | | 10 points | -0.1406 | 1.3993 | 3.7632 | -0.1609 | 1.4266 | 2.8059 | -0.142 | 1.3536 | 2.886 | -0.1618 | 1.4048 | 2.776 |
| | | 20 points | -0.1038 | 1.1287 | 2.3688 | -0.1338 | 1.3359 | 1.7774 | -0.1024 | 1.1626 | 1.5916 | -0.1349 | 1.3467 | 1.7871 |
| | | 30 points | -0.1099 | 1.0197 | 2.7788 | -0.1295 | 1.1649 | 1.8639 | -0.1028 | 0.9656 | 1.8274 | -0.1306 | 1.2318 | 1.9358 |
| | | 1 box | -0.1265 | 3.105 | 0.8339 | -0.0907 | 1.906 | 0.5436 | -0.1297 | 3.2082 | 0.7364 | -0.0913 | 1.9018 | 0.5554 |
| | | 1 box + 1point | -0.1627 | 6.8075 | 1.2629 | -0.1425 | 4.8267 | 0.944 | -0.1727 | 5.613 | 1.1758 | -0.1415 | 4.8046 | 0.9319 |
| | 10-15 | 2 points | -0.1526 | 8.0397 | 8.0968 | -0.285 | 11.9454 | 10.435 | -0.1978 | 8.4389 | 8.718 | -0.2862 | 11.6526 | 10.3761 |
| | | 10 points | -0.2993 | 4.8236 | 10.3039 | -0.3873 | 5.2032 | 9.8519 | -0.3223 | 4.6364 | 9.0634 | -0.3862 | 5.1922 | 9.7825 |
| | | 20 points | -0.2438 | 2.5343 | 7.5686 | -0.3473 | 3.1906 | 7.3885 | -0.2624 | 2.5091 | 6.1821 | -0.3465 | 3.1101 | 7.18 |
| | | 30 points | -0.2567 | 2.6225 | 8.3182 | -0.3452 | 3.2338 | 7.6462 | -0.2642 | 2.4965 | 6.5202 | -0.3444 | 3.1864 | 7.6489 |
| | | 1 box | -0.2048 | 4.9559 | 1.7861 | -0.1774 | 3.3881 | 1.3491 | -0.2122 | 5.1563 | 1.6056 | -0.1781 | 3.4198 | 1.3467 |
| | | 1 box + 1point | -0.2716 | 9.4154 | 3.1145 | -0.2889 | 7.8684 | 2.7739 | -0.2876 | 8.1647 | 2.9253 | -0.2893 | 7.8699 | 2.7976 |

From Table 2, it can be observed that for the same segmented region, as the added randomness gradually increases, the segmentation performance of SAM gradually decreases. For instance, in the segmentation of the whole tumor for T1 modality data, when randomly scaling 0 to 5 pixels,

SAM based on 1 box prompt experienced a decrease in Dice result by 0.0329. When scaling 5 to 10 pixels, the Dice result decreased by 0.1002, and for scaling 10 to 15 pixels, the Dice result decreased by 0.1639. For different segmented regions, SAM's segmentation performance is also slightly affected by random perturbations. For example, in T1 modality data, the Dice for the tumor core decreases by a maximum of 0.3564 (i.e., when adding 10 to 15 pixels of movement based on 10 points), while the Dice for the whole tumor and enhancing tumor do not decrease by more than 0.3 in this scenario.

### 4.1.2 SAM Fine-tuning

SAM consists of three components: the image encoder, prompt encoder, and mask decoder. Due to limited computational resources, we opted to fine-tune a relatively lightweight mask decoder. This involved freezing the parameters of the image encoder and prompt encoder, and during the training process, only adjusting the parameters of the mask decoder. The datasets utilized was the 2D sliced data from BraTS2019, which had been organized in the previous experiments. The data was split into two parts, with one portion serving as the training set and the other as the test set.

It is important to note that in the aforementioned series of experiments, SAM was solely tested on 2D sliced data from BraTS2019 and this process did not involve adjusting SAM parameters. As a result, we can use the 2D sliced data from BraTS2019 again for the fine-tuning training of SAM.

After fine-tuning SAM with a substantial amount of brain tumor datasets, the segmentation performance of SAM, as shown in Table 3, demonstrated a significant improvement. For instance, in the case of whole tumor segmentation under the T1 modality, the Dice score improved from 0.6989 before fine-tuning to 0.8005 after fine-tuning. This highlights SAM's significant potential in downstream tasks. In other words, compared to deploying SAM directly to downstream tasks, fine-tuning SAM with a substantial amount of downstream task-specific data can significantly enhance its performance.

Table 3. Results of SAM on BraTS2019 datasets. Dice, HD, and ASSD are employed for evaluation with SAM.

| Region | Method | Prompts | T1 | | | T1ce | | | T2 | | | FLAIR | | |
|---|---|---|---|---|---|---|---|---|---|---|---|---|---|---|
| | | | Dice↑ | HD↓ | ASSD↓ | Dice↑ | HD↓ | ASSD↓ | Dice↑ | HD↓ | ASSD↓ | Dice↑ | HD↓ | ASSD↓ |
| Whole Tumor | SAM | 1 box | 0.8005 | 17.9673 | 0.6028 | 0.7989 | 17.6185 | 0.6142 | 0.8240 | 17.4714 | 0.5129 | 0.7989 | 17.6185 | 0.6142 |
| Tumor Core | SAM | 1 box | 0.8211 | 14.6222 | 0.4693 | 0.8442 | 14.1880 | 0.3853 | 0.8334 | 14.4117 | 0.4224 | 0.8442 | 14.1839 | 0.3840 |
| Enhancing Tumor | SAM | 1 box | 0.6999 | 14.1850 | 0.7629 | 0.7625 | 12.3332 | 0.5768 | 0.7142 | 13.9397 | 0.7057 | 0.7625 | 12.3332 | 0.5768 |

### 4.2 Future Outlook

Here, we present some potential research directions for SAM:
(i) Medical imaging data is typically three-dimensional, such as MRI, CT, and others. Compared to traditional 2D images, these 3D datasets contain rich spatial contextual information, offering more comprehensive and detailed anatomical information. However, SAM currently only supports 2D images as input. This results in the loss of the original 3D data's spatial contextual information, limiting SAM to independently process each slice without utilizing the positional relationship information between adjacent slices. Therefore, extending SAM to support 3D data is a promising research direction.

ii) The multimodality of medical imaging data is one of its crucial features in clinical and research settings. For instance, BraTS2019 provides four different imaging modalities: T1, T1ce, T2, and FLAIR. Different imaging modalities capture diverse biological and tissue information, providing medical professionals with more comprehensive and accurate patient diagnostic and treatment information. Simultaneously, multimodal data offers researchers a more thorough understanding of biological information, aiding in a deeper exploration of disease mechanisms and individual variations. Currently, SAM processes data independently for different modalities without integrating information from various modalities. In other words, it does not fully utilize the complementary information between different modalities. Therefore, in future work, exploring how SAM can integrate information from different modalities is a research direction.

## 4. CONCLUSIONS

In this study, we evaluated the performance of SAM on brain tumor segmentation and found that: (1) SAM with box prompts performs better than that with point prompts. (2) When more points are provided, SAM's performance further improves (3) If there are too many points, the performance of SAM begins to show a decline. (4) The SAM's performance can further improve when we combine the point prompts and box prompts. (5) The performance of SAM will suffer a certain decrease if we introduce some randomness to the ideal point prompts or box prompts. (6) SAM shows remarkable performance in some special modal datasets, but is unstable in other modal datasets. (7) SAM performs well in segmenting objects with clear boundaries, but its performance is not good for objects with blurry boundaries. (8) Upon fine-tuning SAM with a subset of brain tumor images, a noticeable enhancement in its segmentation performance was observed.